\documentclass[dvipsnames,11pt]{article}
\usepackage{adjustbox}
\usepackage{amssymb}
\usepackage{amsthm}
\usepackage{amsmath}
\usepackage{authblk}
\usepackage{blkarray}
\usepackage{booktabs}
\usepackage{bm}
\usepackage{calc}
\usepackage{caption}
\usepackage{cases}
\usepackage{centernot}
\usepackage[utf8]{inputenc}
\usepackage{dcolumn}
\usepackage{float}
\usepackage[margin = 1in]{geometry}
\usepackage{graphicx}
\usepackage{lineno}
\usepackage{listings}
\usepackage{makecell}
\usepackage{mathtools}
\usepackage[round]{natbib}
\usepackage{pifont}

\usepackage{setspace}
\usepackage{soul}
\usepackage{subcaption}
\usepackage{tabularx}
\usepackage{tgpagella}
\usepackage{verbatim}
\usepackage{xcolor}
\usepackage{xfrac}
\usepackage[style=english]{csquotes}
\MakeOuterQuote{"}
\usepackage{tikz}
\usepackage{marvosym}
\tikzset{mynode/.style={draw, very thick, circle, minimum size=1cm},
    myarrow/.style={very thick, -}}
\newcommand*\patchAmsMathEnvironmentForLineno[1]{%
  \expandafter\let\csname old#1\expandafter\endcsname\csname #1\endcsname
  \expandafter\let\csname oldend#1\expandafter\endcsname\csname end#1\endcsname
  \renewenvironment{#1}%
     {\linenomath\csname old#1\endcsname}%
     {\csname oldend#1\endcsname\endlinenomath}}% 
\newcommand*\patchBothAmsMathEnvironmentsForLineno[1]{%
  \patchAmsMathEnvironmentForLineno{#1}%
  \patchAmsMathEnvironmentForLineno{#1*}}%
\AtBeginDocument{%
\patchBothAmsMathEnvironmentsForLineno{equation}%
\patchBothAmsMathEnvironmentsForLineno{align}%
\patchBothAmsMathEnvironmentsForLineno{flalign}%
\patchBothAmsMathEnvironmentsForLineno{alignat}%
\patchBothAmsMathEnvironmentsForLineno{gather}%
\patchBothAmsMathEnvironmentsForLineno{multline}%
}

\newcommand{\redvoter}{\text{\huge{{\color{red}{\Gentsroom}}}}}
\newcommand{\bluevoter}{\text{\huge{{\color{blue}{\Gentsroom}}}}}

\title{The dynamics of strategic voting: pathways to consensus and gridlock}
\author{ Jonathan Engle\thanks{jae23a@fsu.edu} \, \& Bryce Morsky\thanks{bmorsky@fsu.edu}}
 \affil{ Department of Mathematics, Florida State University, Tallahassee, FL, USA }
\date{\today}
\begin{document}
\maketitle

\begin{abstract}
    The outcomes of democratic elections rest on individuals' decision-making that is driven by their varying preferences and beliefs. Individuals may prefer consensus to gridlock, or gridlock to consensus, and information may be fractured via echo-chambers. To understand the role of these factors in whether or not elections reach consensus, we develop and explore a computational model in which voters have varying party affiliations, preferences, beliefs, and voting strategies. Voters may change their voting strategies either by imitating others or reconsidering their strategy individually. Preferences are orderings of the following election outcomes: a voter's party winning a super-majority, the opposing party winning such a majority, and gridlock. Voters beliefs and decisions are shaped by their social networks, and thus are heterogeneous in the population. We observe a "tipping point" phenomenon wherein the voters' initial strategies and randomness impact whether the minority party voters vote to create gridlock or consensus. A positive feedback loop secures such voters into one behaviour or the other. Consensus is reached by the minority party evolving to prefer consensus, which in turn drives the majority to also prefer consensus due to the influence of social learning. Further, consensus is promoted by an uneven distribution of party affiliation, and undermined when it is even. We also find that a moderate prevalence or strength of echo-chambers can boost consensus, since they can quell voters' desires for gridlock. 
\end{abstract}

\section*{Introduction}
% Paragraph 1) Elections and coordination/consensus/cooperation in politics. But these outcomes depend upon the decision-making of voters (leading to paragraph 2).
Throughout history compromise and cooperation have been crucial in advancing society. Today, as political tensions rise, it becomes increasingly difficult for individuals to cooperate and find common ground to advance legislation \citep{iyengar12}. In recent years, the political landscape has become more hostile and divided as politicians are unable to find common ground \citep{barisione17,vegetti19,birch20,mettler22}. The divide has created more and more extreme ideals and behaviours \citep{huang17}. Under such division, how then do we obtain common ground? Being able to obtain common ground or consensus is heavily dependent on voters' social networks, especially when echo chambers form \citep{cinelli21}. In extreme circumstances, when a polarizing figure is elected, post election voters can be less cooperative, especially with those of other social groups \citep{huang17}. While this is an extreme case, it allows us to understand that while consensus and cooperation are important for societal growth, echo chambers and polarizing leaders can undermine general consensus.

% Paragraph 2) Decision making of voters: strategic voting, zealotry, etc. But these depend on the information voters have (leading to paragraph 3).
On the surface level, voters in elections are faced with choosing the party whose policies most align with their preferences, which can be derived from factors such as income, gender, race, rural or urban living, and many others \citep{wani14,pinto21,mettler22}. But it is not always this trivial, since voters may vote strategically given their beliefs and reasoning about how the rest of the electorate will vote. On the one hand, a voter may prefer legislative action by either party opposed to a gridlock result. This concept directly relates to "do something" politics, where on consensus type issues, voters may prefer that legislative action is taken even if the majority party has opinions that conflicts with their own ideals \citep{egan14}. The severity of an issue also emphasises this drive for legislative action and consensus where a gridlock result is more severe \cite{egan14}. On the other hand, voters may feel that their preferences are inadequately represented by their choices in an election and thus they would prefer gridlock. In either case, a highly important election to a voter can lead to an increase in such strategic voting \citep{myatt07}.

% Paragraph 3) Information bubbles: information gerrymandering (Stewart), how one's social environment affects voting behaviour. We now have several factors affecting election outcomes (leading to paragraph 4).
When deciding on how to vote strategically, voters receive information from a variety of different sources. Historically, information has spread via word of mouth from voters' local social networks of family, friends, and acquaintances. With technological advancements, this local social network has expanded drastically. Many individuals now obtain information via social media platforms, in particular news and political updates \citep{wani14}. The quality of this online information is not always accurate. For example, information absorption online can be faster with lies than truth, and social media platforms can push political agendas \citep{vosoughi18}. Information bubbles and influence networks create a divide among voters by altering their perceived information about the nature of political issues. If such bubbles are adequately "gerrymandered", election outcomes can be altered: even minority parties can have the ability to sway votes in their favour \citep{stewart19}.

% Paragraph 4) Previous mathematical models of elections and voting behaviour. What questions remain that we are aiming to answer in this study? (leading to paragraph 5).
As seen in experiments with human participants concerned with reaching consensus, the decision making of voters can be influenced by their connections to members of the opposing party \citep{stewart19}. In particular, a voter's behaviour can be altered by the "privately observed payoffs" of the neighbors around them \citep{strulovici10}. This effect can be significantly impacted by the presence of political zealots, who vote a party regardless of the opinions of others. If the impact of a minority of zealots is sufficiently large and prevails over multiple elections, then consensus can be prevented leading to prolonged political gridlock \citep{tian18}.

%Paragraph 5) Opinion Dynamics 
The struggle between consensus and gridlock has been studied theoretically through opinion dynamics, where consensus, polarization, or a plurality of options have been observed \citep{hegselmann02, bauso16}. These outcomes are dependent on the initial distribution of prior beliefs as well as the ability and/or willingness to change one's belief. If a voter is willing to change their belief, then it is paramount what type of information is present \citep{acemoglu11}. Opinion dynamics of a voter depend on social influences such as the rate of interaction with a persuader type voter. The persuader type player, if of the same party, can create polarizing dynamics in the voting game \citep{bertotti24}. Polarization can amplify the outcome of consensus if a majority of the population polarizes to the same party or ideal. On the contrary, if the population polarizes evenly, then this effect will lead to gridlock \citep{hegselmann02}. Internal party composition and ideals can also impact the outcome of opinion dynamics. If there is a tight ideological alignment between all members of the same party, then there is an incentive to implement legislation that further aligns with that party undermining the willingness to cooperate with the opposition \citep{nunnari17}.

% Paragraph 6) Our questions and goals and what we have done in this paper. These include: how can consensus be built? How does the voter composition of the population affect the outcome? Do voters converge on zealotry or consensus? What we did to address these questions. Describe briefly our methods without going into the math or specifics.
To understand such a rich interplay among strategic voting, consensus, and varying preferences, we developed an agent-based model of the dynamics of voters' opinions and decisions. It consists of two parties, preferences on consensus or gridlock, and voting strategies including Zealot, Chartist, Consensus-maker, and Gridlocker. The simplest strategy is Zealot in which voters strictly vote for their party regardless of the opinions of others. Nonetheless, the influence of Zealots in the network plays an impact with regards to the observations and decisions made by their neighbours. A Chartist bases their voting on the history of others' past votes with the aim to predict the votes of their neighbours in the future. This is done by using a mechanism from the econophysics literature on the Minority Game \citep{challet97}, which has also included social learning of strategies \citep{morsky23}. Finally, a Consensus-maker votes for the majority party --- they are in essence a "fair weather voter" --- while a Gridlocker votes for the minority party. Gridlockers add an additional barrier to consensus being reached, but do so differently than Zealots. Voters having such strategies are placed on a social network and may imitate the strategies and preferences of their neighbours. We explore how the preferences, strategies, and network of an electorate impact consensus.

\section*{Methods}

We consider the evolution of strategic voting in a two-party system where the outcomes of elections can be categorized as majorities (one party receives greater than half the share of votes) or a super-majorities (one party receives greater than some threshold share of the votes, which is greater than a half). Voters have a party affiliation and "vote" each turn --- which could be a specified period of time or an election --- for either party. Note that these "votes" can be interpreted differently. If the turn is an election, then the "votes" are votes in that election. If the turn is say a week and not an election, then a "vote" is only a declared intention to vote a particular way. Likewise, "winning" on a turn is simply having a (super) majority that turn. For the remainder of the paper, we will not differentiate between these two conceptions of votes, and we will simply speak of turns generically.

Votes are chosen according to voters' strategies, which factor a voter's party affiliation, preferences, and/or information about past elections into determining how to vote. After each turn, voters may change these strategies, but not their party affiliation, by either an individual or a social process. Additionally, voters are embedded in a social network through which they can observe others. We consider random networks for this social network generated by the Erd\"{o}s-Renyi model. Below we discuss the strategies and preferences of voters and how imitation occurs. For specific details of the algorithm for the simulations see Appendix \ref{nummethods}.

\subsection*{Preferences and strategies}

Voters' preferences are an ordering of the following potential outcomes of a turn: their party wins a super-majority, the other party wins a super-majority, or no party wins a super-majority, also denoted as gridlock. We assume that all voters prefer that their party wins a super-majority over the other party winning one. However, they may differ on their ordering of gridlock in their preferences. \emph{Consensus-preferring} voters prefer gridlock least of all while \emph{gridlock-preferring} individuals prefer it most. There is a third preference ordering, which we term \emph{party-preferring}, in which a voter prefers that their party wins a super-majority over gridlock, but that gridlock is reached rather than the other partying winning a super-majority. For each of these preferences, there are strategies employed by voters to determine what vote they should cast.

The first and simplest strategy is Zealot, in which the voter votes for their own party regardless of the votes of their neighbours in the social network. Zealot voters (hereafter called Zealots) may have any three of the preference orderings detailed above. Consensus-preferring Zealots are voters who, though they do not wish gridlock, vote in hope that their party wins a super-majority. On the other hand, gridlock-preferring Zealots do wish for gridlock, and vote their party towards that aim. Finally, party-preferring Zealots most embody party zealotry in that they would vote their party regardless of the votes of all other voters. This is unlike the other two. E.g.,\ consensus-preferring Zealots would be better off voting for the other party if they knew that party alone was near a super-majority.

The next strategies we consider are Consensus-maker and Gridlocker. Consensus-makers vote the same as the majority of their neighbours last turn. This strategy is modelled on a social norm of conformity applied only to direct neighbours of a voter. Thus, Consensus-makers condition their votes only on the past votes of their neighbours rather than of past votes of the entire population. In contrast to Consensus-makers are Gridlockers who vote the opposite of the majority of their neighbours last election. They can be interpreted as voters who have a preference for a third party, which could be niche, extreme, or centrist. Since they are constrained by the two-party system, they vote for gridlock. However, they are still affiliated with one of the parties, a "lesser evil" to which they have the most affinity. We assume that Consensus-makers and Gridlockers may only have the consensus-preferring and gridlock-preferring orderings, respectively.

\begin{table}[!ht]
\begin{center}
\begin{tabular}{cc}
\toprule
Past election winners & Recommended vote \\
\midrule
\tikz\draw[red,fill=red] (0,0) circle (.5ex); \tikz\draw[red,fill=red] (0,0) circle (.5ex); \tikz\draw[red,fill=red] (0,0) circle (.5ex); & \tikz\draw[red,fill=red] (0,0) circle (.5ex); \\
\tikz\draw[red,fill=red] (0,0) circle (.5ex); \tikz\draw[red,fill=red] (0,0) circle (.5ex); \tikz\draw[blue,fill=blue] (0,0) circle (.5ex);  & \tikz\draw[blue,fill=blue] (0,0) circle (.5ex); \\
\tikz\draw[red,fill=red] (0,0) circle (.5ex); \tikz\draw[blue,fill=blue] (0,0) circle (.5ex); \tikz\draw[red,fill=red] (0,0) circle (.5ex);  & \tikz\draw[red,fill=red] (0,0) circle (.5ex); \\
\tikz\draw[red,fill=red] (0,0) circle (.5ex); \tikz\draw[blue,fill=blue] (0,0) circle (.5ex); \tikz\draw[blue,fill=blue] (0,0) circle (.5ex);  & \tikz\draw[red,fill=red] (0,0) circle (.5ex); \\
\tikz\draw[blue,fill=blue] (0,0) circle (.5ex); \tikz\draw[red,fill=red] (0,0) circle (.5ex); \tikz\draw[red,fill=red] (0,0) circle (.5ex);  & \tikz\draw[red,fill=red] (0,0) circle (.5ex); \\
\tikz\draw[blue,fill=blue] (0,0) circle (.5ex); \tikz\draw[red,fill=red] (0,0) circle (.5ex); \tikz\draw[blue,fill=blue] (0,0) circle (.5ex);  & \tikz\draw[blue,fill=blue] (0,0) circle (.5ex); \\
\tikz\draw[blue,fill=blue] (0,0) circle (.5ex); \tikz\draw[blue,fill=blue] (0,0) circle (.5ex); \tikz\draw[red,fill=red] (0,0) circle (.5ex);  & \tikz\draw[blue,fill=blue] (0,0) circle (.5ex); \\
\tikz\draw[blue,fill=blue] (0,0) circle (.5ex); \tikz\draw[blue,fill=blue] (0,0) circle (.5ex); \tikz\draw[blue,fill=blue] (0,0) circle (.5ex);  & \tikz\draw[red,fill=red] (0,0) circle (.5ex); \\
\bottomrule
\end{tabular}
\end{center}
\caption[]{A strategy table for a Chartist that conditions recommended votes on the outcomes of the last three elections. \tikz\draw[red,fill=red] (0,0) circle (.5ex); and \tikz\draw[blue,fill=blue] (0,0) circle (.5ex); are the two different parties. The outcomes of the last three elections are represented by a string of the winning vote on turn $k-3$, $k-2$, and $k-1$ for current turn $k$ in that order.} \label{table:strategy_table}
\end{table}

The fourth and final strategy is called Chartist, which uses the past history of neighbours' votes to predict the likely majority next turn. Specifically, Chartists use strategy tables, which have previously been used in models of the Minority Game \citep{challet97}, to determine their votes. Strategy tables allow voters to condition their votes on the last $3$ turns. Each Chartist has $2$ strategy tables, which provide a recommended vote for each possible combination of past majority votes of their neighbours. Each strategy table thus has $2^3$ recommended votes. Table \ref{table:strategy_table} represents a strategy table with strings of the past three majority votes along with recommended votes for the current turn. During a turn, all strategy tables earn payoffs as if they were followed and as a function of the strategy table's recommendation, the voter's affiliation, and the vote that turn. Each turn, Chartists follow the recommendation of their strategy table with the highest payoff: i.e.,\ the strategy table that has historically been most accurate. Let $\pi(i,j,m)$ be the payoff of a strategy table where $i$ is the voter's party affiliation, $j$ was the strategy table's recommendation given the outcomes of the last $3$ turns, and $m$ is the majority vote of the current turn. Then, for a consensus-preferring Chartist affiliated with party $i$, $\pi(i,i,i) > \pi(i,j,j) > \pi(i,i,j) > \pi(i,j,i)$ for $i \neq j$. The strategy table earns the highest payoff if it recommended to vote for the voter's party and that was the majority. The next highest payoff is when the other party won a majority and the strategy table recommended to vote for that party. Since, such a vote strengthens the win of the majority party (even though the voter isn't affiliated with it). The next payoff is where the strategy table recommended the voter vote against the party that went on to win a majority, but for the party with which the voter is affiliated. The strategy table predicted the majority incorrectly, but at least recommended voting for the voter's party. The lowest payoff a strategy table receives is when the recommended vote does not match the winning party and is not for the voter's party. Intuitively, the strategy table will be promoted/demoted if it recommends for/against the majority party, since such a party is nearer to (or at) a super-majority. By similar arguments, the ordering of payoffs for gridlock-preferring Chartists  affiliated with party $i$ is: $\pi(i,j,i) > \pi(i,i,j) > \pi(i,i,i) > \pi(i,j,j)$ for $i \neq j$. Note that Chartists cannot be party-preferring, since rationally a voter with such a preference should always vote their party and not use any strategy tables.

\begin{table*}[!ht]
\begin{center}
\begin{tabular}{lll}
Strategy name & Voting behaviour & Election preferences \\
\midrule
Chartist & use strategy tables to determine vote & consensus or gridlock \\
Consensus-maker & vote for the majority last election & consensus \\
Gridlocker & vote for the minority last election & gridlock \\
Zealot & always vote for their party & consensus, gridlock, or party
\end{tabular}
\end{center}
\caption{Voters strategies and preferences for the election outcome.} \label{table:types_of_voters}
\end{table*}

\sloppy
In total, there are fourteen types of voters, seven for each party affiliation: consensus-preferring Chartists, gridlock-preferring Chartists, Consensus-makers, Gridlockers, consensus-preferring Zealots, gridlock-preferring Zealots, and party-preferring Zealots. An overall summary of voters' strategies and election preferences are given in Table \ref{table:types_of_voters}.

\subsection*{Imitation dynamics}

After every turn, each voter will consider changing its strategy by selecting one of its neighbours to learn from. Since voters of the same party affiliation receive the same payoff regardless of their individual votes, we measure the "fitness" differential of voters by considering whether or not their votes were for the majority, how much they each prefer that majority, and whether or not the voters have the same party affiliation (see Appendix \ref{nummethods} for details). As an example, consider a population of only Consensus-makers. The probability that a row Consensus-maker imitates a neighbouring column Consensus-maker given that the red party received the majority vote is presented in the following matrix:
\begin{align}
    &P(\tikz\draw[red,fill=red] (0,0) circle (.5ex);) = \bordermatrix{
        ~ & \redvoter \tikz\draw[red,fill=red] (0,0) circle (.5ex); & \redvoter \tikz\draw[blue,fill=blue] (0,0) circle (.5ex); & \bluevoter \tikz\draw[red,fill=red] (0,0) circle (.5ex); & \bluevoter \tikz\draw[blue,fill=blue] (0,0) circle (.5ex); \cr
        \redvoter \tikz\draw[red,fill=red] (0,0) circle (.5ex); & 1/2 & 0 & \phi/2 & 0 \cr
        \redvoter \tikz\draw[blue,fill=blue] (0,0) circle (.5ex); & 1 & 1/2 & \phi & \phi/2 \cr
        \bluevoter \tikz\draw[red,fill=red] (0,0) circle (.5ex); & \phi/2 & \phi/4 & 1/2 & 1/4 \cr
        \bluevoter \tikz\draw[blue,fill=blue] (0,0) circle (.5ex); & 3\phi/4 & \phi/2 & 3/4 & 1/2 \cr
    }, \label{eqn:imitation} \\
    &\redvoter / \bluevoter \text{ red/blue voter},
    \tikz\draw[red,fill=red] (0,0) circle (.5ex);/\tikz\draw[blue,fill=blue] (0,0) circle (.5ex); \text{ voted red/blue}. \notag
\end{align}
The probability that a red Consensus-maker who voted blue imitates a blue Consensus-maker who voted red is $\phi$ (second row third column). The parameter $\phi \in [0,1]$ is applied when the voters have different party affiliations. The smaller $\phi$, the less voters will imitate neighbours with a different party affiliation and thus, relatively, the more they will imitate those of the same affiliation. Similar homophilic imitation dynamics have been explored in evolutionary game theory using the replicator equation \citep{morsky17}. The imitation probabilities when blue received the majority vote are similar, and the complete set of payoffs given differing neighbours is supplied in Appendix \ref{nummethods}. Note here that voters are not necessarily imitating by comparing their fitness with the fitness of their neighbours, but on whether or not their neighbour's vote would have benefited them more than the vote their own strategy determined given their party affiliation.

\section*{Results}

\begin{figure}[ht!]
    \centering
    \includegraphics[width=0.65\linewidth]{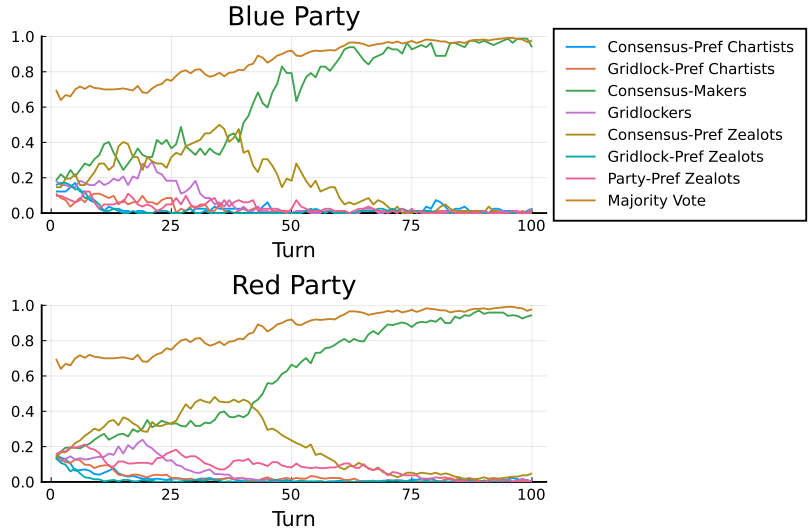}
    \caption{A representative time series where consensus is reached. Here, $\beta=0.8$, $\phi=1$, and the mean node degree is $7$.}
    \label{fig:ts_beta_0.8}
\end{figure}

\begin{figure}[ht!]
    \centering
    \includegraphics[width=0.65\linewidth]{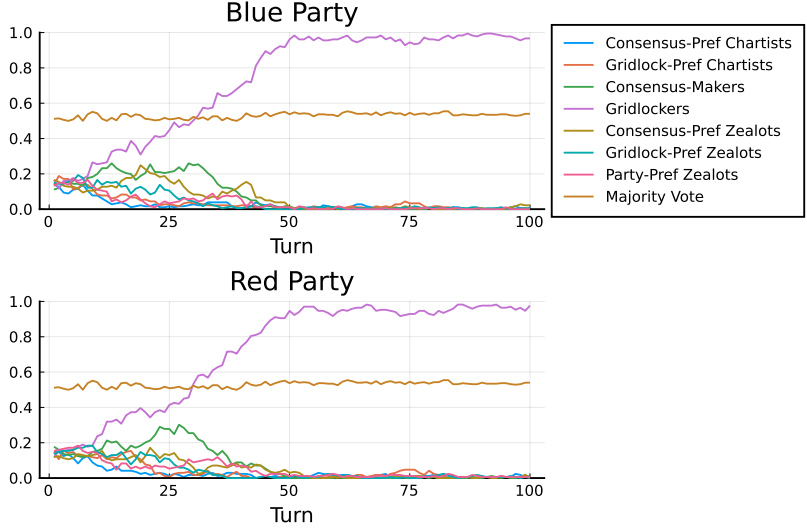}
    \caption{A representative time series where consensus is not reached and instead Gridlockers dominate. Here, $\beta=0.5$, $\phi=1$, and the mean node degree is $7$.}
    \label{fig:ts_beta_0.5}
\end{figure}

\begin{figure}[ht!]
    \centering
    \includegraphics[width=0.65\linewidth]{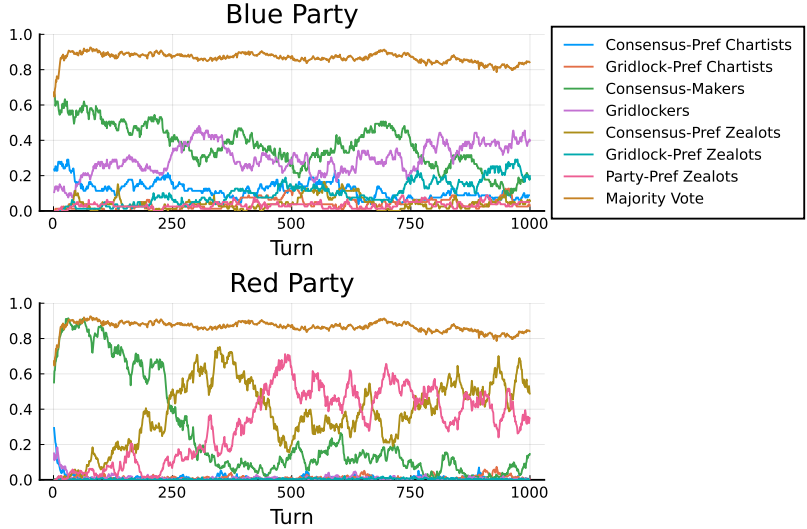}
    \caption{A representative time series of long term fluctuations in strategies where consensus is reached. Here, $\beta=0.8$, $\phi=0$, and the mean node degree is $7$.}
    \label{fig:ts_oscillations}
\end{figure}

In our simulations for the following results, we consider a network of $350$ voters who are each assigned to a party with probability $\beta$, our party bias parameter. The network is then constructed and each voter is randomly connected to other voters. To begin with, we consider the case where the probability of connecting voters in the initial network construction is the same regardless of party affiliation. We plot heatmaps of the average degree of consensus at the end of these simulations for a variety of initial conditions. The $x$-axis in these figures details the initial proportions, in equal measure, of consensus-preferring Chartists and Consensus-makers (but not consensus-preferring Zealots), i.e.\ consensus-preferring non-Zealots. Similarly, the $y$-axis represents the proportion of the electorate that are initially gridlock-preferring Chartists or Gridlockers (i.e.\ gridlock-preferring non-Zealots). The remaining initial population is Zealots (in equal measure of each preference ordering). For each of these maps we average over $50$ realizations with each game containing $250$ turns (see Appendix \ref{nummethods} for further simulation details). For Figures \ref{fig:ts_beta_0.5}, \ref{fig:ts_beta_0.8}, and \ref{fig:ts_oscillations}, we take a deeper look at the strategy dynamics and composition of note worthy cases by analyzing the time series. In the final figures, we consider cases where the formation of the social network is dependent on the party affiliation of voters.

% Fig 1) Heatmaps for p=q=0.02. Vary beta and phi. Heatmaps represent initial conditions. x-axis: Consensus-makers + consensus-preferring Chartists. y-axis: Gridlockers + gridlock preferring Chartists. z: Zealots (includes all different preferences). x+y+z=1. 4 panels.
\begin{figure}[ht!]
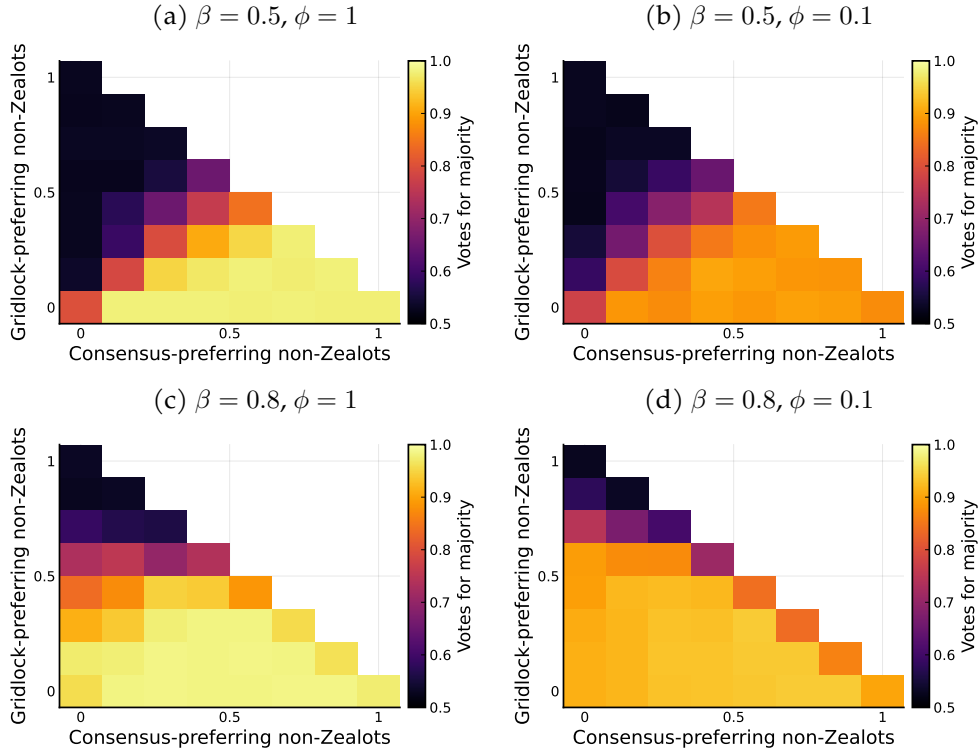

\captionsetup[subfigure]{justification=centering}
    \centering
    \begin{subfigure}[]{0.4\columnwidth}
        \caption{$\beta=0.5$, $\phi=1$}\label{fig:heatmaps_sparse_a}
        \includegraphics[width=\textwidth]{Figuresjpeg/Figure1a.jpeg}
    \end{subfigure}
    \begin{subfigure}[]{0.4\columnwidth}
        \caption{$\beta=0.5$, $\phi=0.1$}\label{fig:heatmaps_sparse_b}
        \includegraphics[width=\textwidth]{Figuresjpeg/Figure1b.jpeg}
    \end{subfigure} \\
    \begin{subfigure}[]{0.4\columnwidth}
        \caption{$\beta=0.8$, $\phi=1$}\label{fig:heatmaps_sparse_c}
        \includegraphics[width=\textwidth]{Figuresjpeg/Figure1c.jpeg}
    \end{subfigure}
    \begin{subfigure}[]{0.4\columnwidth}
        \caption{$\beta=0.8$, $\phi=0.1$}\label{fig:heatmaps_sparse_d}
        \includegraphics[width=\textwidth]{Figuresjpeg/Figure1d.jpeg}
    \end{subfigure}
    \caption{Heat maps for mean degree $7$.}
    \label{fig:heatmaps_sparse}
\end{figure}

Figure \ref{fig:heatmaps_sparse} depicts the results for varying degrees of party bias and homophily with mean node degree of $7$. We see that the system is bi-stable: either a high or low degree of consensus is reached. If, initially, there are many consensus-preferring Chartists and Consensus-makers, then a high degree of consensus is reached. However, if there are more gridlock-preferring Chartists and Gridlockers initially, then consensus is low. Bias in the party affiliation of the electorate results in a larger basin of attraction to consensus than no bias (Figures \ref{fig:heatmaps_sparse_c} and \ref{fig:heatmaps_sparse_d} vs.\ \ref{fig:heatmaps_sparse_a} and \ref{fig:heatmaps_sparse_b}). A high prevalence of the majority party promotes consensus among the minority. A low $\phi$ (high degree of relative homophily) reduces the influence of majority party members on the minority, and thus the overall vote for consensus. Additionally, a unique phenomenon occurs for the edge cases of our heat maps when there is a high bias and initially few Zealots (Figures \ref{fig:heatmaps_sparse_c} and \ref{fig:heatmaps_sparse_d}). We observe that the degree of consensus is less than that of a more diverse strategy set showing that a degree of zealotry in the population is desirable to drive the population to a super-majority.

% Fig 2) Heatmaps for p=q=0.2. Vary beta and phi. Heatmaps represent initial conditions. x-axis: Consensus-makers + consensus-preferring Chartists. y-axis: Gridlockers + gridlock preferring Chartists. z: Zealots (includes all different preferences). x+y+z=1. 4 panels.
\begin{figure}[ht!]
\captionsetup[subfigure]{justification=centering}
    \centering
    \begin{subfigure}[]{0.4\columnwidth}
        \caption{$\beta=0.5$, $\phi=1$}
        \includegraphics[width=\textwidth]{Figuresjpeg/Figure2a.jpeg}
    \end{subfigure}
    \begin{subfigure}[]{0.4\columnwidth}
        \caption{$\beta=0.5$, $\phi=0.1$}
        \includegraphics[width=\textwidth]{Figuresjpeg/Figure2b.jpeg}
    \end{subfigure} \\
    \begin{subfigure}[]{0.4\columnwidth}
        \caption{$\beta=0.8$, $\phi=1$}\label{fig:heatmaps_dense_c}
        \includegraphics[width=\textwidth]{Figuresjpeg/Figure2c.jpeg}
    \end{subfigure}
    \begin{subfigure}[]{0.4\columnwidth}
        \caption{$\beta=0.8$, $\phi=0.1$}
        \includegraphics[width=\textwidth]{Figuresjpeg/Figure2d.jpeg}
    \end{subfigure}
    \caption{Heatmaps for mean degree $70$.}
    \label{fig:heatmaps_dense}
\end{figure}

Figure \ref{fig:heatmaps_dense} depicts the case where the mean degree is $70$. We observe many of the same qualitative differences between varying parameter values as when the mean degree is $7$. Additionally, for $\beta=0.8$, notice that this higher mean degree extends the basis of attraction relative to mean degree of $7$, which can be seen by comparing Figures \ref{fig:heatmaps_sparse_c} and \ref{fig:heatmaps_dense_c}. The higher the mean degree of the network, the more individuals are impacted by social learning. The large mean degree allows for voters to learn from a larger portion of the network and thus have a more accurate  view of the opinions of other voters. This fact aids convergence to consensus, particularly when there is a bias in the party affiliation of voters.

% Figs 3-5) Representative time series for several interesting cases.
% Fig 3) beta=0.5, gridlock and little consensus

To understand the trajectories to consensus and gridlock found in the previous heatmaps, we plot representative time series of these scenarios in Figures \ref{fig:ts_beta_0.8} and \ref{fig:ts_beta_0.5}, respectively. For Figure \ref{fig:ts_beta_0.8}, bias in party affiliation of the red party is high ($\beta=0.8$), there is no homophily ($\phi=1$), and the mean node degree is $7$. With initial conditions of all strategy types being equal, an interesting dynamic emerges. In these skewed type cases (by party), both the majority (the red party) and minority (the blue party) obtain a majority strategy of Consensus-makers. It is perhaps unsurprising in this case that the minority party adopts the Consensus-maker strategy, since there is an abundance of voters affiliated with the majority party, and that consensus on the red party is reached. Additionally, one may expect that members of the majority party would be indifferent between zealotry and consensus-like strategies. However, since there is a high prevalence of the Consensus-maker strategy among the minority and that the majority is not homophilic, the majority adopts the Consensus-maker strategy from the minority.

% Fig 4) beta=0.5, consensus

Figure \ref{fig:ts_beta_0.5} is a representative time series where consensus is not reached, and depicts how the dynamics lead to a high prevalence of the Gridlocker strategy among voters of either party affiliation. Here, $\beta=0.5$, $\phi=1$, and the mean node degree is $7$. Notice how initially, while both parties struggle to gain control of the majority, the majority strategy is consensus-preferring Zealots. A similar result was observed in the gridlock dynamics of \cite{stewart19}. After approximately $30$ turns, both parties switch to the Gridlocker strategy. Because voters' social networks are likely to be composed of voters of both parties in equal amounts, the observed majority reaches approximately $50\%$, which gives Gridlockers the highest payoff and thus promotes the spread of that strategy in the population.

% Fig 5) phi=0, zealotry and gridlock, instability

In Figure \ref{fig:ts_oscillations}, we consider a special case where there is large bias in party affiliation ($\beta=0.8$) and there are no cross-party imitation dynamics ($\phi=0$): i.e., there is no probability that a voter will imitate the strategy of members of the other party. Additionally, we initialize the simulation with a high distribution of Consensus-makers and consensus-preferring Chartists with a small distribution of Gridlockers, similar to the boundary cases of Figure \ref{fig:heatmaps_sparse}. Consensus-makers initially increase in frequency among the majority red party affiliated voters, but eventually fluctuate along with consensus-preferring Zealots and party-preferring Zealots. Once a significant number of majority voters have adopted Zealot strategies, which occurs just before turn $250$, Gridlocker proliferates among blue party affiliated voters, which persists for a long period of time. Zealots and Consensus-makers also persist among the majority. Even though Gridlockers are abundant among the minority, consensus is still high due to the party bias and presence of Consensus-makers among the minority. However, the level of consensus is less than when cross-party imitation is present.

% Fig 6) varying p and q
\begin{figure}[ht!]
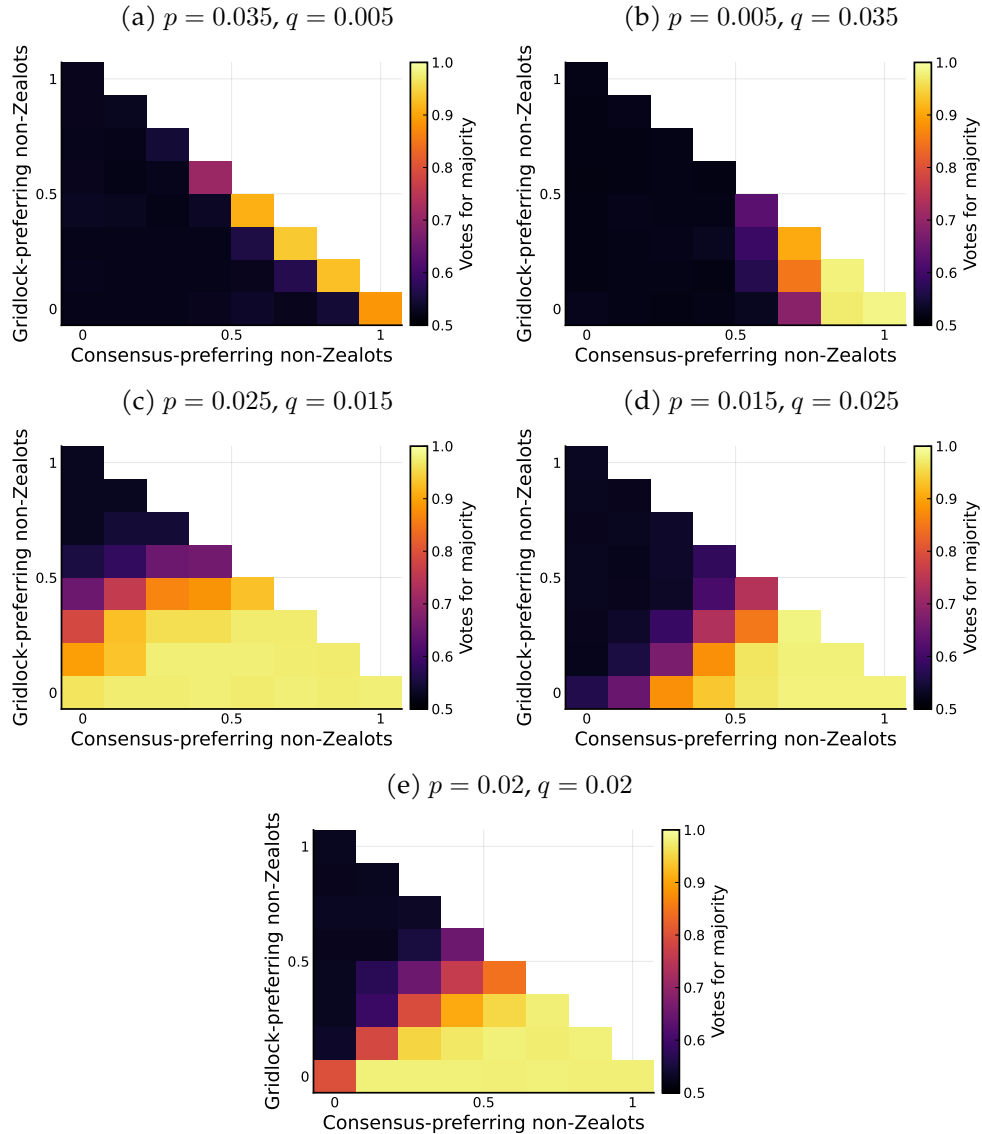

\captionsetup[subfigure]{justification=centering}
    \centering
    \begin{subfigure}[]{0.4\columnwidth}
        \caption{$p=0.035$, $q=0.005$}\label{fig:heatmaps_p_not_q_a}
        \includegraphics[width=\textwidth]{Figuresjpeg/Figure6a.jpeg}
    \end{subfigure}
    \begin{subfigure}[]{0.4\columnwidth}
        \caption{$p=0.005$, $q=0.035$}\label{fig:heatmaps_p_not_q_b}
        \includegraphics[width=\textwidth]{Figuresjpeg/Figure6b.jpeg}
    \end{subfigure} \\
    \begin{subfigure}[]{0.4\columnwidth}
        \caption{$p=0.025$, $q=0.015$}\label{fig:heatmaps_p_not_q_c}
        \includegraphics[width=\textwidth]{Figuresjpeg/Figure6c.jpeg}
    \end{subfigure}
    \begin{subfigure}[]{0.4\columnwidth}
        \caption{$p=0.015$, $q=0.025$}\label{fig:heatmaps_p_not_q_d}
        \includegraphics[width=\textwidth]{Figuresjpeg/Figure6d.jpeg}
    \end{subfigure} \\
    \begin{subfigure}[]{0.4\columnwidth}
        \caption{$p=0.02$, $q=0.02$}\label{fig:heatmaps_p_not_q_e}
        \includegraphics[width=\textwidth]{Figuresjpeg/Figure6e.jpeg}
    \end{subfigure}
    \caption{Heatmaps for varying values of $p$ and $q$, where $p$ is the probability of a voter connecting to voters affiliated with their party and $q$ is the probability of a voter connecting to a voter of the party. Here $\beta=0.5$, $\phi=1$, and the mean degree is $7$.}
    \label{fig:heatmaps_p_not_q}
\end{figure}

Next, we introduce party biased connections between voters, where a voter is connected to voters of their or the opposing party with different probabilities, $p$ and $q$, respectively. These results are depicted in Figure \ref{fig:heatmaps_p_not_q} along with the heatmap from Figure \ref{fig:heatmaps_sparse_a} for comparison. For a moderate degree of party assortativity, consensus is improved due to Gridlockers earning low payoffs (Figure \ref{fig:heatmaps_p_not_q_c}). Gridlockers generally earn high payoffs when networks are well-mixed with voters of both parties and as these voters struggle for control. When voters are more likely to be connected to those that share their party affiliation, their social networks contain less party diversity, which drives Gridlockers' payoffs down. However, if party assortativity is too high, consensus is undermined as depicted in Figure \ref{fig:heatmaps_p_not_q_a}. In this case, gridlock is reached not by a high prevalence of Gridlockers, but by high levels of zealotry.

If a voter is more likely to connect with voters of the opposing party, as shown in seen in Figures \ref{fig:heatmaps_p_not_q_b} and \ref{fig:heatmaps_p_not_q_d}, then overall consensus is decreased. These mixed echo chambers require higher initial proportions of consensus-preferring non-Zealots to reach consensus. Note that in these figures we have only shown the results for $\beta=0.5$ and $\phi=1$, since varying these parameters have a similar impact as discussed earlier. Except, contrary to the previous figures, the impact of the homophily parameter has minimal impact when $q$ is low, since there is already a low degree of connection between voters of opposing parties (results not shown here).

% Fig 7) time series for different p and q
\begin{figure}[ht!]
    \centering
    \includegraphics[width=0.65\linewidth]{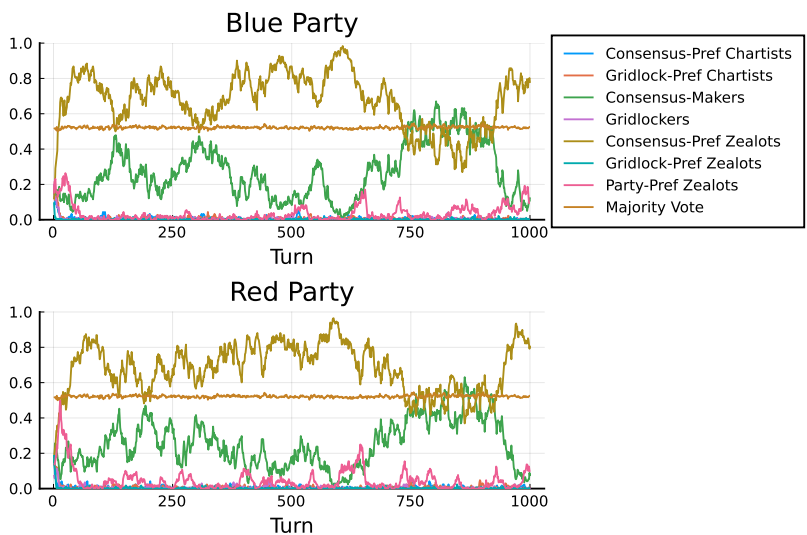}
    \caption{A representative time series of long term fluctuations in strategies. Here, $p=0.035$ is the probability of a voter connecting to voters affiliated with their party and $q=0.005$ is the probability of a voter connecting to a voter of the other party. The mean degree for all voters being $7$.}
    \label{fig:ts_pnotq}
\end{figure}

Plotting the time series for a simulation in which $p=0.035$ and $q=0.005$, it can be seen see that Zealots become more prominent for both parties (Figure \ref{fig:ts_pnotq}). Comparing these results to Figure \ref{fig:ts_oscillations}, we observe similar dynamics for the majority party of Figure \ref{fig:ts_oscillations} and both parties in Figure \ref{fig:ts_pnotq}: there is an inverse relationship in the oscillations between Consensus-makers and Zealots. However, in Figure \ref{fig:ts_pnotq}, the party-preferring Zealots are less prominent than in Figure \ref{fig:ts_oscillations}. In the case of Figure \ref{fig:ts_oscillations}, the majority red party voters are indifferent between the party-preferring Zealot and consensus-preferring Zealot strategies due to strong in-group effects. In both scenarios, echo chambers --- either driven by low cross-party imitation rates or high assortativity by party --- act as a barrier to social learning from voters of the other party. A key difference between these two scenarios, however, is that the high party bias in Figure \ref{fig:ts_oscillations} does lead to high levels of consensus, while there is only gridlock in Figure \ref{fig:ts_pnotq}.

% Fig 8) large party echo chambers
\begin{figure}[ht!]
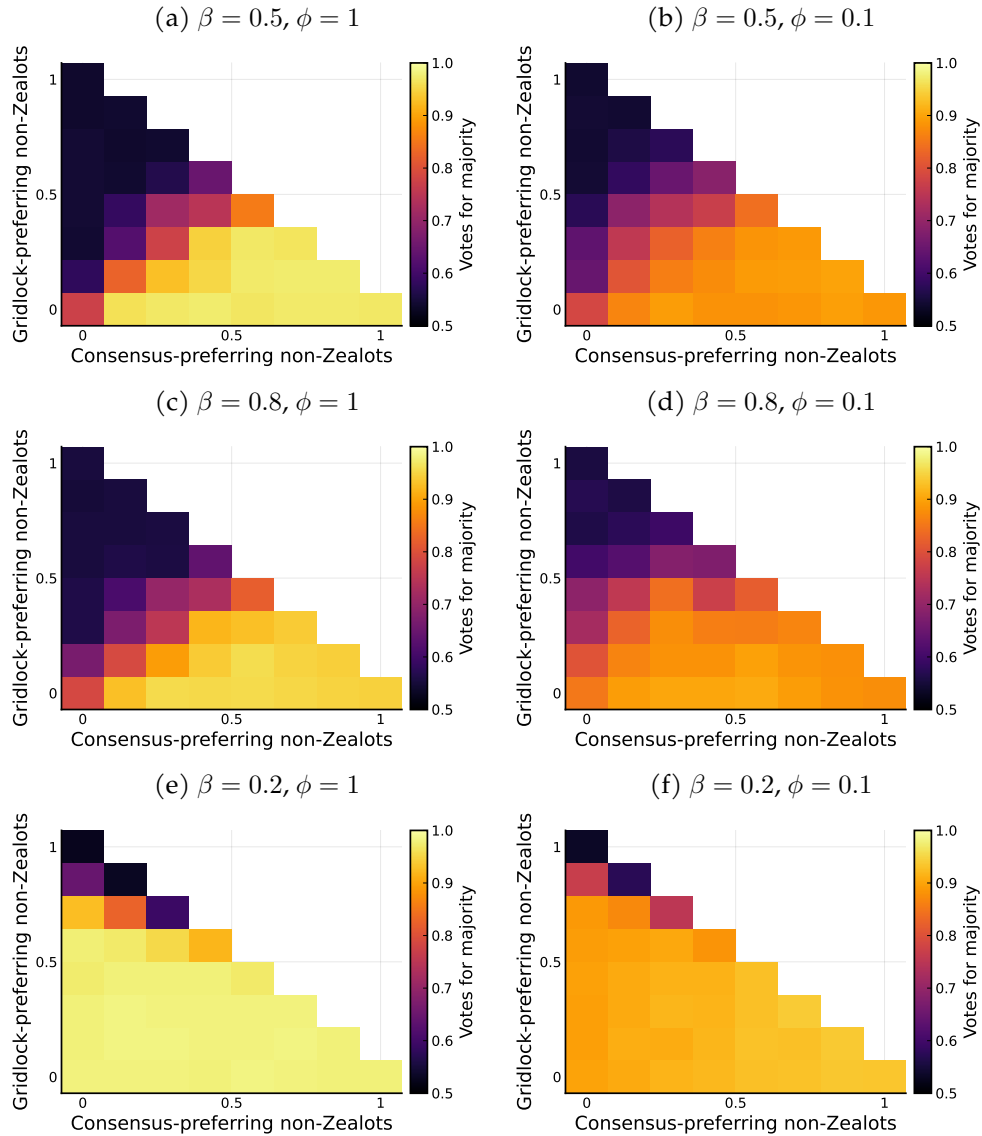

\captionsetup[subfigure]{justification=centering}
    \centering
    \begin{subfigure}[]{0.4\columnwidth}
        \caption{$\beta=0.5$, $\phi=1$}
        \includegraphics[width=\textwidth]{Figuresjpeg/Figure8a.jpeg}
    \end{subfigure}
    \begin{subfigure}[]{0.4\columnwidth}
        \caption{$\beta=0.5$, $\phi=0.1$}
        \includegraphics[width=\textwidth]{Figuresjpeg/Figure8b.jpeg}
    \end{subfigure} \\
    \begin{subfigure}[]{0.4\columnwidth}
        \caption{$\beta=0.8$, $\phi=1$}\label{fig:heatmaps_pblue_qred_c}
        \includegraphics[width=\textwidth]{Figuresjpeg/Figure8c.jpeg}
    \end{subfigure}
    \begin{subfigure}[]{0.4\columnwidth}
        \caption{$\beta=0.8$, $\phi=0.1$}
        \includegraphics[width=\textwidth]{Figuresjpeg/Figure8d.jpeg}
    \end{subfigure}\\
    \begin{subfigure}[]{0.4\columnwidth}
        \caption{$\beta=0.2$, $\phi=1$}\label{fig:heatmaps_pblue_qred_e}
        \includegraphics[width=\textwidth]{Figuresjpeg/Figure8e.jpeg}
    \end{subfigure}
    \begin{subfigure}[]{0.4\columnwidth}
        \caption{$\beta=0.2$, $\phi=0.1$}
        \includegraphics[width=\textwidth]{Figuresjpeg/Figure8f.jpeg}
    \end{subfigure}
    \caption{Heatmaps for when blue party affiliated voters are more connected than red ones. $p_\text{blue}=0.02$ is the probability that a blue party voter selected under the Erd\"{o}s-R\'{e}nyi process will be connected to another individual (even if the neighbour is red). And $q_\text{red}=0.01$ is the probability a red party voter selected to connect with another voter will be connected regardless of party.}
    \label{fig:heatmaps_pblue_qred}
\end{figure}

Finally, we introduce another party preferring connection where the blue party affiliated voter is more likely to connect with another voter regardless of party affiliation than a red party affiliated voter, i.e.\ blue voters are better connected than red voters. This high connection rate increases the overall influence of the blue party on the network. This counteracts the effect of a red-biased electorate. Figure \ref{fig:heatmaps_pblue_qred_c} shows that even with a higher party bias in favour of the red party, there is still a smaller degree of consensus compared to the even connections case (Figure \ref{fig:heatmaps_sparse}). Similarly, for a high party bias in favour of the blue party (Figure \ref{fig:heatmaps_pblue_qred_e}), the overall degree of consensus is significantly higher and gridlock states are only possible for high initial conditions of gridlock-preferring non-Zealots. Notice that a high imitation rate as well as a high probability of a voter being connected to a blue voter promotes consensus across Figure \ref{fig:heatmaps_pblue_qred}. Additionally, as the imitation rate goes down due to homophily (right panels vs. the left) then the overall impact of the blue voters is dampened as there is less probability of red voters imitating them. This in turn, decreases the overall degree of consensus.

\section*{Discussion}

Here we have developed a model of two-party elections and voting behaviour in which voters have varying party preferences and strategies. We considered three preference orderings: consensus, gridlock, and party. From a game theoretical view, the consensus-preferring voters are essentially playing a coordination game such as an $N$-player Battle of the Sexes \citep{wang22}. Coordination games have also been studied under the context of how gerrymandered information can undermine a super-majority \citep{stewart19}. Our results extend that of \cite{stewart19} by incorporating other preferences and strategies. We show that the addition of Gridlockers and Zealots can undermine consensus. More specifically, initial conditions, randomness, and the degree of homophily play a critical role in determining whether the minority party voters vote to create gridlock or consensus and in turn whether or not the majority also adopts consensus preferences. Though we have framed our results in terms of promoting consensus and a strong majority, we should note that gridlock may not necessary be bad. It has both costs \citep{teter12} and benefits to a republic \citep{gerhardt13}. On the one hand, for example, gridlock can protect minorities from a unchecked majority. On the other hand, it can impede progress.
 
One of our main results is a theoretically validation of the effect of "insecure majorities" \citep{lee16}. Such majorities are characterized as tenuous and liable to be lost in the next election. This has the effect of undermining cooperation between political parties as they each vie for a majority in the coming election. In our case, the balance of party affiliation is what increases election competitiveness, since there is no clear majority. We thus find that when the electorate is well-balanced in their party affiliation, majorities are often smaller and voters tend to neither vote for consensus nor prefer it. This effect is only exacerbated if the electorate is heavily information gerrymandered or highly homophilic, since voters are then primarily influenced by those who are like-minded. In contrast, unbalanced party affiliation leads to more consensus making from the minority party and thus larger majorities.

To broadly explore our model, we considered a variety of strategy compositions of the population. Such compositions are likely highly context dependent in reality. Some studies have shown low levels of strategic voting and a high degree of "wasted" votes \citep{felsenthal85,nunez16,heath22}. For example, a 2017 survey of voters from the 2017 Uttar Pradesh, India showed little strategic voting \citep{heath22}. This was due to voters' overestimating the chance their party would win. The few voters who do vote strategically tend to base their votes on polling and decide their vote late in the election \citep{felsenthal85}. Yet, strategic voting has been observed in other elections. For example, in the U.S.\ Senate election in Florida in 2010, voters readily switched from the Democrat candidate Kendrick Meek to the Independent Charlie Crist on evidence that Crist was more likely to defeat the Republican Marco Rubio \citep{mckee13}. Strategic voting was also present in the 2008 U.S.\ primary elections, though was more prevalent later in the primary process \citep{hillygus14}. Additionally, the impact of strategic voting can be context dependent. For example, strategic voting had a greater impact on the U.K.\ general elections of 1997 than of 2001 even though the rates of strategic voting were similar \citep{herrmann16}. Another example is voter complacency where voters tend to revert "default" voting behaviours where there is minimal strategy involved, which can lead to inefficiencies \citep{tal15}. An increased prevalence of strategic voters reduces the incentives for others to vote strategically as well \cite{myatt07}.

There are several limitations of this study coming from our assumptions that could be relaxed in future work. For one, we assumed an undirected influence network. Though this effectively models friends, neighbours, and acquaintances interacting and sharing political opinions, it does not model one directional influence. Such asymmetrical influence may arise from the nature of the interaction (e.g.\ online posting of opinions), or differing attributes of individuals (e.g.\ social statuses, persuasiveness, or desire to persuade). Additionally, we did not model party evolution or the internal dynamics of parties, either by voters switching parties or parties changing their positions to attract voters. The political system may also be highly driven by its history. For example, election victories of one's party tends to boost approval relative to compromise, while a gridlock result damages party allegiance \cite{flynn16}. Future models could incorporate such phenomena and dynamics as well as additional parties. In particular, the strategies explored here may perform substantially differently in multi-party parliamentary elections.

% \subsection*{Statements and Declarations}
% The authors have no relevant financial or non-financial interests to disclose.

\subsection*{Code and Data Availability}
Code to run the simulations is available at github.com/bmorsky/voting-game.

\appendix
\section{Simulation details} \label{nummethods}
The following describes the simulations of elections and behaviour of each voter in the network. In each election, we consider $350$ voters with each Chartist containing $2$ strategy tables. To plot the heatmaps, we vary different initial conditions for consensus-preferring non-Zealots and gridlock-preferring non-Zealots, and average each cell over $50$ realizations that are run for $250$ turns each. For the time series figures, initial conditions vary to illustrate key cases and dynamics in the heat maps. We then generate a random Erd\"{o}s-Reyni graph by connecting a voter to another voter of the same party with probability $p$ and of the other party with probability $q$. After constructing this network, we initialize the first turn with each player intending to vote for their respective parties. Payoffs for the players are then computed. The payoff function for consensus-preferring voters is:
\begin{equation}
    \pi_c(i,j,m)=\begin{cases}
        1 &\text{ if }i=j=m,\\
        -1 &\text{ if } i \neq j\neq m,\\
        1/2 &\text{ if } i\neq j= m,\\
        -1/2 &\text{ if } i=j \neq m,
    \end{cases}
\end{equation}
where $i$ is their vote, $j$ is their party and $m$ is the local majority vote among their neighbours. Similarly, the payoff functions for gridlock-preferring ($\pi_g$) and party-preferring ($\pi_p$) voters are:
\begin{align}
    &\pi_g(i,j,m)=\begin{cases}
        -1/2 &\text{ if }i=j=m,\\
        1/2 &\text{ if } i \neq j\neq m,\\
        -1 &\text{ if } i\neq j= m,\\
        1 &\text{ if } i=j \neq m,
    \end{cases} \\
    &\pi_p(i,j,m)=\begin{cases}
        1 &\text{ if }i=j=m,\\
        -1 &\text{ if } i=j\neq m.\\
    \end{cases}
\end{align}
Note that $\pi_p$ only considers the cases where the majority is or is not the same as the voters' party and their vote, since these voters always vote their party and do not value consensus.

Since the maximum and minimum payoff differentials are $2$ and $-2$, respectively, we normalize the payoff differential by adding two and dividing by four. Then, we define the probability that a party $i$ affiliated voter who voted for $j$ imitates a party $k$ voter who voted for $l$ when party $m$ has the majority is:
\begin{equation}
    P_{ij\to kl}(m) = \frac{\phi_{ik}}{4}(2+\pi(i,k,m)-\pi(i,j,m)).
\end{equation}
We assume that $\phi_{ii} = 1 \geq \phi_{ij} = \phi \in [0,1]$ for $i \neq j$. From this equation we can find the elements of the matrix of Equation \ref{eqn:imitation}. Essentially, the focal voter considers what payoffs they would have earned if they had the same strategy as the voter they are considering imitating and compare that to what they have earned. Then, this is modulated by whether they are on the same party or not.

Imitation then occurs over several rounds. Additionally, each voter every turn may individually learn a new strategy with probability $\mu=0.01$. When this occurs, they choose a random new strategy. In summary, the algorithm for an election is then given as:
\begin{enumerate}
    \item Initialize the game by assigning voters a node, party, and initial strategy.
    \item Compute payoffs for all voters.
    \item Each voter then may socially or individually learn a new strategy.
    \item Voters then choose whether to profess to vote for their party or the other party.
    \item Repeat steps $2$-$4$ for $T$ turns.
\end{enumerate}

\bibliography{voting}
\bibliographystyle{apsr}

\end{document}